\newcommand{\sla}[1]{#1\!\!\!/}
\newcommand{\ba}{\begin{eqnarray}}
\newcommand{\ea}{\end{eqnarray}}

\documentclass[aps,preprint,showpacs,showkeys]{revtex4}
\usepackage{graphicx}
\usepackage{amssymb}

\begin{document}

\title{FERMION PRODUCTION IN STRONG MAGNETIC FIELD AND ITS  ASTROPHYSICAL IMPLICATIONS}

\author{Hyun Kyu Lee\footnote{hyunkyu@hanyang.ac.kr} and Yongsung Yoon\footnote{cem@hanyang.ac.kr}}

\affiliation{Department of Physics, Hanyang University, Seoul 133-791, Korea}

\begin{abstract}

We calculate the effective potential of a strong magnetic field induced by fermions with anomalous magnetic moments
which couple to the electromagnetic field  in the form of the Pauli interaction.  For a uniform magnetic field,  we
find the explicit form of  the effective potential. It is found that the non-vanishing imaginary part develops for a
magnetic field stronger than a critical field  and has  a quartic form which is quite different from the exponential
form of the Schwinger process. We also consider a linear magnetic field configuration as an example of inhomogeneous
magnetic fields. We find that the imaginary part of the effective potential is nonzero even below the critical field
and shows an exponentially decreasing behavior with respect to the inverse of the magnetic field gradient, which is the
non-perturbative characteristics analogous to the Schwinger process. These results  imply the instability of the strong
magnetic field to produce fermion pairs as a purely magnetic effect. The possible applications to the astrophysical
phenomena with strong magnetic field are also discussed.

\end{abstract}



\maketitle

\section{Introduction}

Recent observations of the explosive phenomena like magnetars and gamma ray bursts indicate the possibility of very
strong magnetic field around the compact objects\cite{wt,grb} which are considered to be responsible for powering the
explosive events. The inferred field strength are larger than $\sim 10^{15}$ G, which is stronger than the conventional
critical value $ 4 \times 10^{13}$ G. While most of the astrophysical environment is electrically neutral except the
very narrow region near the surfaces of compact objects\cite{usov}, the rotation of the magnetic field induces an
electric field $\sim v_{rotation} B$. For $B \sim 10^{15}$ G, the induced electric field is estimated to be  order of
$\sim 10^{17}$ V/cm.  Apparently the induced electric field strength  is much larger than the critical field strength
for electro-positron pair creation via Schwinger process\cite{schwinger}. However the invariant $ B^2- E^2>0$ for this
configuration indicates that the electromagnetic field configuration is dominated by the magnetic field.  It has been
known that  the production of the minimally interacting fermion is not possible in the magnetically dominated
configuration\cite{dunn}. The pair production of minimally interacting particles is  a purely electric effect. Hence
the strong magnetic field configuration in the astrophysical environment is considered to be stable against the
particle creation\cite{duncan}.

While the minimal coupling derived by the local gauge invariance is of fundamental nature, there appear also
non-minimal couplings as well in the form of effective theory. Pauli introduced a non-minimal coupling of spin-1/2
particles with electromagnetic fields, which can be interpreted as an effective interaction of fermions with an
anomalous magnetic moment\cite{pauli,lavrov,ho}. For the neutral fermions with non-vanishing magnetic moments, it is
the Pauli interaction through which  the electromagnetic interaction can be probed. It is interesting to note that the
inhomogeneity of the magnetic field, which couples directly to the magnetic dipole moment through Pauli interaction,
plays a similar role analogous to the electric field for charged particles with the minimal coupling. The possibility
of production of the neutral fermions in a purely magnetic field configuration  with spatial inhomogeneity has been
demonstrated in 2+1 dimension\cite{lin}, and  the production rate in 3+1 dimension  has been calculated explicitly for
the magnetic field with a spatial inhomogeneity of a critical value\cite{leeyoon}. Recently the possibility of particle
creations even in the uniform magnetic field\cite{ly} has been demonstrated provided the field strength is stronger
than the critical value, which is the ratio of the fermion mass to its anomalous magnetic moment. This implies that the
magnetic field configuration becomes unstable and the pair creation is possible as a purely magnetic effect. In section
2, the Pauli interaction is introduced and discussed in comparison with the minimal coupling.  The details of the
calculations of effective potentials for the uniform and the linear magnetic field configuration will be discussed in
section 3 and 4 respectively.  The results are summarized and their astrophysical implications are discussed in section
5.

\section{Pauli interaction}

The Dirac Lagrangian of a neutral fermion with Pauli interaction is given by
\begin{equation}
{\cal L} =
\bar{\psi}(\sla{p}+\frac{\mu}{2}\sigma^{\mu\nu}F_{\mu\nu}-m)\psi,\label{pauli}
\end{equation}
where $\sigma^{\mu\nu}=\frac{i}{2}[\gamma^{\mu},\gamma^{\nu}],
~~g_{\mu\nu}=(+,-,-,-)$.  $\mu$ in the Pauli term measures the
magnitude of the  magnetic moment of fermion. The corresponding
Hamiltonian is given by
\begin{equation}
H_{pauli} = \vec{\alpha}\cdot (\vec{p}-i\mu
\beta\vec{E})+\beta(m-\mu \vec{\sigma}\cdot\vec{B}),
\label{hamiltonian}
\end{equation}
where $ \sigma^{i}=\frac{1}{2}\epsilon^{ijk}\sigma^{jk}$. It is
also known that the Hamiltonian for a charged  fermion with
minimal coupling is given by
\begin{equation}
H_{dirac} = \vec{\alpha}\cdot (\vec{p}-e\vec{A})+\beta(m-\beta e
\phi). \label{dhamiltonian}
\end{equation}
One  can note a kind of duality between $ \vec{B}$ of Pauli Hamiltonian and $\phi$ of Dirac Hamiltonian modulo $\gamma$
matrices. Let us consider the  Pauli Hamiltonian with pure magnetic field, $\vec{E}=0$, and the Dirac Hamiltonian with
pure electric field, $\vec{A}=0$.  Then, for a static limit, $\vec{p}=0$, we get
\begin{equation}
H_{pauli} \rightarrow \beta(m-\mu \vec{\sigma}\cdot\vec{B}), ~~ H_{dirac} \rightarrow \beta(m-\beta e \phi).
\label{hamiltonian0}
\end{equation}
In this simple static limit, a particle state is considered to be assigned with
$\beta=+1$ and antiparticle state with $\beta=-1$. Then the energy difference between the particle and antiparticle
state is given by
\begin{equation}
\Delta H_{pauli} \rightarrow 2(m-\mu\vec{\sigma}\cdot\vec{B}), \label{hamiltonian0d}
\end{equation}
for the Pauli interaction and
\begin{equation}
\Delta H_{dirac} \rightarrow 2m, \label{dhamiltonian0d}
\end{equation}
for the minimal
interaction.   One can see that there is a level crossing with Pauli interaction for a strong enough  magnetic field.
For a uniform magnetic field, the energy eigenvalues of the Hamiltonian Eq.($\ref{hamiltonian}$) are given by
\begin{equation}
E=\pm \sqrt{p_{l}^{2}+(\sqrt{m^{2}+p_{t}^{2}} \pm \mu B)^{2}},
\label{energy}
\end{equation}
where $p_{l}$ and $p_{t}$ are respectively the longitudinal and the transversal momentum to the magnetic field
direction. One can see that for a magnetic field stronger than the critical field $B_{c}=\frac{m}{\mu }$, the ground
state with $p_{l}=p_{t}=0$ crosses  zero energy state.
 Theoretically the particle creation is known to be associated
with  level crossing\cite{gj}. Analogously in this example one can easily guess the pair creation for $B > B_c$: the
state occupied the negative energy sea becomes a particle state and the vacant positive energy state plunges into the
negative sea to make an antiparticle state.
 This indicates the possible instability of the  magnetic field configuration.  In section 3, we will show explicitly that the
imaginary part of the effective potential develops exactly when $B = B_c$. On the other hand, the energy eigenvalues of
minimally interacting charged fermions are
\begin{equation}
E=\pm \sqrt{p_{l}^{2}+m^{2}+|e|B(2n+1-{\rm sgn}(e)\hat{s})},
\label{energy2}
\end{equation}
where $n=0,1,2,\ldots$, and $\hat{s}=\pm 1$ is the spin projection along the magnetic field\cite{khalilov}. It should
be pointed out that the ground state energy is independent of the external magnetic field and moreover no state plunges
into the zero energy state even for a strong magnetic field.  The observation that  particle production of minimally
interacting fermions is impossible in pure magnetic fields can be attributed to this finite energy gap.

However, for a charged particle with minimal coupling,  this energy gap, $2m$,  can be overcome  by the virtual work
done by the electric field up to the Compton wave length. This process has been known as Schwinger
process\cite{schwinger} with  a pair creation rate similar to the tunnelling process\cite{field}. The approximate
duality between $ \vec{B}$ of Pauli Hamiltonian and $\phi$ of Dirac Hamiltonian indicates that the inhomogeneity of the
magnetic field  manifested in the form of $\vec{\nabla}\times \vec{B}$, which couples directly to the magnetic dipole
moment through Pauli interaction, plays a similar role analogous to the electric field for a charged particle with the
minimal coupling. It is interesting to note that this duality is exact in 2+1 dimension\cite{lin}.   Hence we can
expect fermion pair creation even for $B < B_c$.  Denoting the spatial derivative as $|B'|$, we can expect the particle
creation for strong enough magnetic field such that the virtual work along the Compton wavelength, $\lambda = 1/m$, is
comparable to the mass:
\begin{equation}
\mu ~ |B'| ~ \lambda \sim  m.
\end{equation}
In analogy to the Schwinger process  the production rate is
expected to have a following form: \begin{equation}
 w  \sim m^4 \exp^{- {\rm  constant} \times
m^2/|\mu  B'|}.\label{expo}
\end{equation}
The details of the calculation of effective potential for the linear magnetic field configuration will be discussed in
section 4.

\section{Effective potential for the uniform magnetic field}

In general, the effective potential, $ V_{\rm eff}(A)$,  for a background electromagnetic vector potential, $A_{\mu}$,
can be obtained by integrating out the fermion:
\begin{eqnarray}
-i \int d^{4}x V_{\rm eff}(A[x]) = \int d^{4}x <x|
tr\ln\{(\sla{p}+\frac{\mu}{2}\sigma^{\mu\nu}F_{\mu\nu}-m)\frac{1}{\sla{p}-m}\}|x>,
\end{eqnarray}
where $F_{\mu\nu}=\partial_{\mu}A_{\nu} - \partial_{\nu}A_{\mu}$,
and $tr$ denotes the trace over Dirac algebra.
 The decay probability
of the background magnetic field into the neutral fermions is related to the imaginary part of the effective potential
$V_{\rm eff}(A)$,
\begin{equation}
P = 1-|e^{i \int d^{4}x  V_{\rm eff}(A[x])}|^{2} = 1-e^{-2\Im \int d^{3}x dt V_{\rm eff}(A[x])}.
\end{equation}
That is, the twice of the imaginary part of the effective potential $V_{\rm eff}(A[x])$ is the fermion production rate
per unit volume\cite{field}: $w(x)= 2\Im (V_{\rm eff}(A[x]))$ for small probabilities.

For a uniform magnetic field configuration such that $\vec{B}=B \hat{z}$, the integral form of the effective potential
is obtained as\cite{leeyoon}
\begin{equation}
V_{\rm eff}=-\frac{(\mu B)^{2}}{4\pi^{2}}
\int^{\infty}_{0}\frac{ds}{s^{2}} [
i\int^{1}_{0}d\xi(1-\xi)e^{i(\mu B)^{2}
\xi^{2}s}-\frac{i}{2}+\frac{(\mu B)^{2}s}{12} ] e^{-im^{2}s}.
\label{VC}
\end{equation}
This integration can be done explicitly. Introducing dimensionless parameters, $t=m^{2}s$ and $\beta=\frac{|\mu
B|}{m}$, the imaginary part of the effective potential Eq.($\ref{VC}$) can be  written as
\begin{eqnarray}
\Im(V_{\rm eff})&=& -\frac{m^{4}\beta^{2}}{8\pi}
\int^{1}_{0}d\xi(1-\xi)[1 - \beta^{2}\xi^{2} -|1-\beta^{2}\xi^{2}|
]. \label{im}
\end{eqnarray}
One can easily see that for a magnetic field weaker than the critical field, $\beta \leq 1$, the integration
Eq.($\ref{im}$) vanishes. It can be also verified by a contour integration\cite{leeyoon}. Therefore, one can see that
the uniform magnetic fields weaker than the critical field are stable as expected.

However, for a magnetic field stronger than the critical field, $\beta > 1$, the imaginary part of the effective
potential does not vanish, but gives a positive value:
\begin{equation}
\Im(V_{\rm eff})=\frac{1}{48\pi}(|\mu  B|-m)^{3}(|\mu B|+3m).
\label{im2}
\end{equation}
This is exactly what is  expected that the pair creation is associated with  the level crossing of positive and
negative energy states. It is interesting to note that the development of the imaginary part associated with a level
crossing has been also demonstrated in the different contexts, for example in the work of Graham and Jaffe\cite{gj}.
 Therefore,  the uniform magnetic fields stronger than the critical field are unstable, and reduce
their strengths producing the fermion pairs. This implies that
there is an upper bound for stable magnetic field strengths
determined by the particle whose ratio of the mass to the magnetic
moment, $\frac{m}{\mu}$, is the least among the fermions which
couple to electromagnetic fields through the Pauli interaction.
Pauli interaction in 2+1 dimension shows quite different physics
compared that in 3+1 dimension. The development of the
non-vanishing imaginary part of the effective potential for
uniform magnetic fields stronger that the critical field strength
does not happen in 2+1 dimension. It is because the energy gap
between the particle and the antiparticle states is $2m$, which is
independent of magnetic field strength even in the presence of the
Pauli interaction in 2+1 dimension.

For the completeness, the real part of the effective potential
Eq.($\ref{VC}$) has been  also calculated explicitly as shown in
FIG. 1. For a weak field, $\beta \ll 1$, $\Re(V_{\rm eff})$
approximates to $\frac{(\mu B)^{6}}{240\pi^{2}m^{2}}$, and for the
critical field, $\beta=1$, $\Re(V_{\rm
eff})=(96\ln2-65)\frac{m^{4}}{288\pi^{2}}$. The real and imaginary
parts of the effective potential with respect to the magnetic field
strength are shown together in FIG.1 in the unit of
$\frac{m^{4}}{48\pi^2}$.

\begin{figure}[htp]
\begin{center}
\includegraphics[height=2in,width=2.4in]{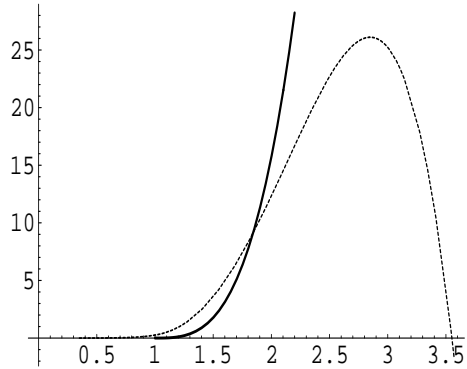}
\caption{Effective potential of the uniform magnetic field B for neutral fermions with a magnetic moment: vertical axis
is $V_{\rm eff}$ in the unit of $\frac{m^{4}}{48\pi^2}$ (the solid line is for the imaginary part and the dashed line
is for the real part), horizontal axis is $\beta(=|\mu B|/m$).}
\end{center}
\end{figure}

 It is also interesting to see how the
instability due to the Pauli interaction is  affected when the
minimal coupling is turned on in addition to the Pauli
interaction. Let us consider an effective Lagrangian, which might
describe a fermion endowed with  a non-vanishing electric charge
$e$ and as well as with a magnetic dipole moment $\mu$, given by
\begin{equation}
{\cal L} =
\bar{\psi}(\sla{p}-e\sla{A}+\frac{\mu}{2}\sigma^{\mu\nu}F_{\mu\nu}-m)\psi.
\label{pauli_e}
\end{equation}
In this work, for the simplicity we consider  the electric charge $e$ and the anomalous magnetic dipole moment $\mu$ as
two independent couplings.  It is straightforward to show that the imaginary part is not affected by the addition of
the independent minimal coupling\cite{ly}.  Hence we can see that for the fermion described by  Eq.(\ref{pauli_e})
 the minimal coupling does not affect  the instability due to the
magnetic moment through the Pauli interaction.

In QED, however,  $e$ and  $\mu$ are not  independent parameters. $\mu$ could be identified as the Schwinger's
anomalous magnetic moment $\mu_a = \frac{\alpha}{2\pi}\frac{e}{2m} $,  which comes from the 1-loop radiative
corrections\cite{field}. The full calculation of the QED radiative correction for strong magnetic
fields\cite{radiative} shows that the Pauli term description of the Schwinger's anomalous magnetic moment is valid only
for weak magnetic fields such that $B \ll m^{2}/e$, which is much smaller than the critical field for the anomalous
magnetic moment defined by $B_c= m/\mu_a$.  It is found that there is no level crossing associated with the pair
creation and the pure magnetic configuration is stable with minimally interacting charged fermions\cite{duncan}.

\section{Production rate for a linear magnetic field}

In the previous section, we observe that  the anomalous magnetic
moment induces an instability of uniform magnetic field for $B \geq
B_c$.  However, the instability due to the anomalous magnetic moment
even for $B \leq B_c$  occurs provided that the field gradient $B'$
is strong enough to overcome the energy gap as discussed in Section
2.

We consider a static linear magnetic field configuration with a constant gradient along an orthogonal direction to the
magnetic field in 3+1 dimension\cite{leeyoon}. We take $\hat{z}$-direction as the magnetic field direction with a
constant gradient along $\hat{x}$-direction, $\vec{B}=B(x)\hat{z}$, such that
\begin{equation}
F_{12}=B(x)=B_{0}+B'x=B'\widetilde{x},~~ (\widetilde{x} = x_{*}+x,
~~x_{*}=\frac{B_{0}}{B'}),
\end{equation}
where the field gradient $B'$ is a non-zero constant. After
lengthy but straightforward calculations we get the effective
potential  given by
\begin{eqnarray}
V_{\rm eff} &=& -\frac{(\mu B)^{2}}{4\pi^{2}} \int^{\infty}_{0}\frac{ds}{s^{2}}\{ i\sqrt{\mu B's\coth(\mu B's)}
\int^{1}_{0} d\xi(1-\xi) e^{i\frac{(\mu B)^{2}}{\mu B' }\xi^{2}\tanh(\mu B's)}-\frac{i}{2}
 \label{VG}\\ &~&
 +\frac{(\mu B)^{2}s}{12} \}e^{-im^{2}s} +\frac{1}{8\pi^{2}}\int^{\infty}_{0} \frac{ds}{s^{3}} \{ (\mu B's\coth (\mu
B's))^{3/2}-1-\frac{(\mu B's)^{2}}{2} \} e^{-im^{2}s}. \nonumber
\end{eqnarray}
The divergent contributions at $s=0$ are removed by adding local
counter terms of $(\mu B)^{2}$, and $(\mu B)^{4}$ in the first
integral. This implies the renormalization of the magnetic moment
$\mu$ to the measured value and the coupling of $(\mu B)^{4}$ to
zero presumably. An additional divergent contribution at $s=0$ is
removed by adding a local counter term of $(\mu B')^{2}$ in the
second integral.

The effective potential for the uniform field configuration
Eq.($\ref{VC}$) can be also obtained by putting  $\mu B' =0$ in
Eq.($\ref{VG}$).

Introducing dimensionless parameters defined as  $t = s m^2$ and $\beta'= \frac{|\mu B'|}{m^{2}}$, the production rate
density $w(x)$ in the unit of the fermion mass is finally given by
\begin{eqnarray}
w(x) =
&-&\frac{2m^{4}}{4\pi^{2}\lambda\kappa}\int^{\infty}_{0}\frac{dv}{v^{2}}
\{ \sqrt{v\coth v} F(\frac{\lambda}{\kappa}\tanh v,\lambda v)
-\frac{1}{2}\cos\lambda v -\frac{\lambda v}{12\kappa}\sin\lambda v
\} \nonumber
\\
&-&\frac{m^{4}}{4\pi^{2}\lambda^{2}}\int^{\infty}_{0}\frac{dv}{v^{3}}\{(v
\coth v)^{3/2}-1-\frac{v^{2}}{2} \}\sin \lambda v, \label{w}
\end{eqnarray}
where
\begin{eqnarray}
F(a,b) &\equiv& \int^{1}_{0}d\xi (1-\xi) \cos(a\xi^{2}-b).
\nonumber
\end{eqnarray}

For a linear magnetic field configuration, the magnetic field
variation at the scale of the Compton wavelength of the fermion is
$|B'|/m$. Therefore, the background magnetic field weaker than
$|B'|/m$ is unphysical. That is, the spatial gradient of the
background magnetic field $|B'|$ should be smaller enough than
$m|B|$, $\beta' < \beta$. Since the integration Eq.($\ref{w}$) has
essential singularities along the imaginary axis, we have not
succeeded to get an analytic expression of the integrations in
Eqs.(\ref{VG},\ref{w}), but can obtain the numerical results, in
which the the physical requirement, $\beta' <  \beta$, turns out
to be the necessary condition to make the production rate density
$w(x)$ positive.

As seen in FIG.2 and FIG.3, provided that the field gradient is extremely large, fermion pair creation is possible for
a magnetic field weaker than the critical field. The particle production rate shows an exponentially decreasing
behavior with respect to the inverse of the field gradient for a fixed strength $|B| < B_{c}$ in the form of
Eq.(\ref{expo}), which is the characteristics of the non-perturbative process. This can be understood as a quantum
tunnelling through an energy gap $\sim 2(m-\mu B)$ of a particle with the help of quantum energy fluctuation $ \sim \mu
|B'|/m$ due to the inhomogeneous magnetic field coupled to the anomalous magnetic moment through Pauli interaction.
This is similar to the Schwinger process of electron-positron pair creation overcoming the mass gap $2m$ in the  strong
electric field with the help of quantum energy fluctuation $|e E|/m$, where the production rate is decreasing
exponentially\cite{field}, $w \sim e^{- {\rm constant} \times m^2/|e E|}$.

However, for magnetic fields stronger than the critical field, it
turns out that the spatial inhomogeneity of magnetic fields does
not help fermion production, but reduces the production rate as
seen in FIG.2. It is found that the production rate density
Eq.($\ref{w}$) approaches to
$\frac{m^{4}}{24\pi}(\beta-1)^{3}(\beta+3)$, which is the
production rate density for a uniform magnetic field from
Eq.(\ref{im2}), as $\beta' \rightarrow 0$.

\begin{figure}[htp]
\begin{center}
 \includegraphics[height=2in,width=2.4in]{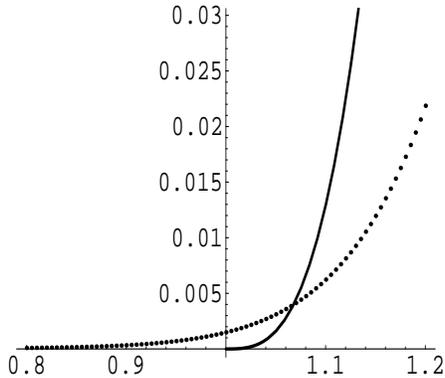}
 \caption{Production rate density $w$ in the unit of $\frac{m^{4}}{4\pi^{2}}$ for $\beta'=0.1$(dotted line) and
 $\beta'=0$(solid line) with varying $\beta$.}
\end{center}
\end{figure}

\begin{figure}[htp]
\begin{center}
\includegraphics[height=2in,width=2.4in]{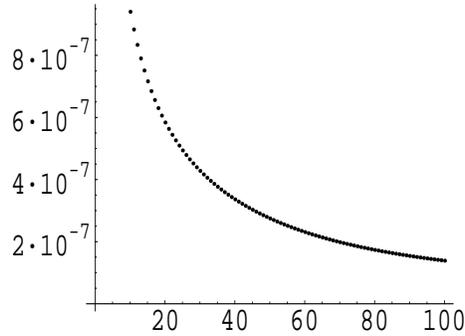}
\caption{Production rate density $w$ in the unit of
$\frac{m^{4}}{4\pi^{2}}$ for $\beta=0.1$ with varying $1/\beta'.$}
\end{center}
\end{figure}

\section{Discussion}
In this paper, we discuss the possibility  of particle creation in the strong magnetic field as a magnetic effect in
contrast to the Schwinger process for creating  a charged pairs as an electric effect.  It has been well known that
pair creation of minimally interacting fermions is not possible in magnetically dominant
configurations\cite{schwinger,dunn,duncan}. However, it is found that in a purely magnetic field configuration the
instability of the magnetic field also occurs due to the anomalous magnetic moment of spin-1/2 fermions through Pauli
interaction\cite{leeyoon,ly}.
 The instability  induced by the neutral fermion with  magnetic moment has been
demonstrated explicitly by showing the emergence of imaginary parts of the effective potential either  when the field
strength becomes stronger than the critical field $B_c=\frac{m}{\mu }$  or when the gradient of the magnetic field is
strong enough $B' \sim m B_c$.
 For a uniform magnetic field $B \geq B_c$,
we get  a simple analytical form of the production rate density $w$ given by $\frac{m^{4}}{24\pi}(\frac{|\mu
B|}{m}-1)^{3}(\frac{|\mu B|}{m}+3)$. We also demonstrate that the production rate for the inhomogeneous magnetic field
is of order $m^4$ for $B \sim B_c$. These results  imply the instability of the strong magnetic field to produce
fermion pairs as a purely magnetic effect.

 Recent observations of  the
explosive astrophysical phenomena like Soft Gamma Repeaters(SGRs),
Anomalous X-ray Pulsars(AXPs)\cite{wt} and Gamma Ray
Bursts(GRBs)\cite{grb} indicate that  the extraordinarily strong
magnetic fields ($\geq 10^{15}$G) can be  expected in the vicinity
of the compact objects, neutron stars and black holes, which are
supposed to be located at the center of the activities.   Since the
astrophysical environment is most likely neutral, the environments
are considered to be magnetically dominant.  It has been argued that
the magnetized vacuum is stable against the electron-positron pair
creation and the only instability might be due to the monopole pair
creation near the Planck scale\cite{duncan} which leads to the
critical field strength $B_c \sim 10^{23} - 10^{53}$G, as far as
particles with non-vanishing electric or magnetic charges are
considered.  However provided that  there is a neutral fermion for
which the interaction with an electromagnetic field is described by
Pauli interaction, Eq.(\ref{pauli}),  it is very interesting to
discuss whether the magnetic vacuum  instability near the critical
field  might have any interesting implication on the astrophysical
phenomena dominated by strong magnetic field.

As an example, neutrinos  are electrically neutral but are known to have  nonzero-mass from the neutrino oscillation
observations\cite{nuosc}. So far  there is no experimental evidence for the magnetic moment of the neutrinos but one
can not simply rule out the magnetic moment just because of the electrical neutrality. In a minimal extension of the
standard model\cite{fujikawa} to incorporate the neutrino mass the magnetic moment of neutrino is calculated to be
$\sim 10^{-20} \mu_B$ if we take the neutrino mass as a cosmological bound on the neutrino mass\cite{cosmology},
$m_{\nu}\sim 10^{-1}$ eV.  Beyond standard model calculations\cite{ng,ch,majorana} yield also a wide range of neutrino
magnetic moment up to the current laboratory upper limit\cite{nulimit}, $\mu_{\nu} < 10^{-10} \mu_B$.
 Then the lower
limit of the  critical field strength is given by
$B_c=\mu_{\nu}/m_{\nu} \geq 10^{17}$G when we take  the neutrino
magnetic moment as a current upper limit.   It is interesting to
note that  the critical field strength $B_c \sim 10^{17}$G is not
so inconsistent with  $10^{15}$G inferred from SGRs, AXPs and GRBs
in the sense that it can be attained assuming a specific dynamo in
the neutron star\cite{dynamo}.   The production rate of the
neutrino pair due to the magnetic vacuum instability for the
critical field is proportional to $m_{\nu}^4$ and is given as $ w
\sim 10^{31} /{\rm m^3 s}$.   The neutrino luminosity can be
estimated as \ba L_{\nu} &\sim& w \times  m_{\nu} \times R_{NS}^3
\sim 10^{30} {\rm erg/s}. \label{lumin}\ea  However it turns out
to be much smaller than the X-ray luminosity of magnetars, $L_X
\sim 10^{35}$erg/s such that  any appreciable effects due to the
magnetic vacuum instability via neutrino pair production may  be
hardly observed from the magnetars.

Basically for a particle with mass $m$ and magnetic moment $\mu$,
the pair production rate is $\sim m^4$ near the critical field
strength $B_c= m/\mu$.  The characteristic time scale is given by
$\tau_B \sim B_c^2/m^5$.   With $R_B^3$ as an effective volume of
the magnetosphere, the luminosity can be estimated as $L \sim m^5
R_B^3$.  Provided  $B_c \sim B_{magnetar}$ and $R_B \sim 10^{4}$m
as those of a magnetar,  a larger luminosity for observation can
be easily obtained  with  a massive particle but  the magnetic
moment should be increasing to keep the critical field strength,
which constrains significantly models for particles involved.
Hence for an interesting observable effects due to the magnetic
vacuum instability to be discussed there should be a physically
well-motivated effective theory which admits the Pauli interaction
for particles with mass and magnetic moment appropriate to the
environment, perhaps most likely related to the astrophysical
phenomena, which produces a strong magnetic field up to the
critical field strength.

This work was supported by grant No. (R01-2006-000-10651-0) from
the Basic Research Program of the Korea Science \& Engineering
Foundation.

\end{document}